\documentclass[times,trackchanges]{aastex631}
\usepackage{soul}
\usepackage{comment}

\received{xxxx, 2024}
\revised{xxxx, 2024}
\accepted{March 13, 2025}
\submitjournal{ApJ}

\shorttitle{proper motion and natal kick in AT2019wey}
\shortauthors{Cui, Jiang, An et al.}
\graphicspath{{./}{figures/}}

\newcommand{\masyr}{mas\,yr$^{-1}$}
\newcommand{\kms}{km\,s$^{-1}$}

\begin{document}

\title{Proper Motion and Natal Kick in the Galactic Black Hole X-ray Binary AT2019wey}

\correspondingauthor{Lang Cui}
\email{cuilang@xao.ac.cn}

\author[0000-0003-0721-5509]{Lang Cui}
\affiliation{Xinjiang Astronomical Observatory, CAS, 150 Science-1 Street, Urumqi 830011, China}
\affiliation{Key Laboratory of Radio Astronomy and Technology, CAS, A20 Datun Road, Chaoyang District, Beijing, 100101, China}
\affiliation{Xinjiang Key Laboratory of Radio Astrophysics, 150 Science 1-Street, Urumqi 830011, China}

\author[0000-0003-3166-5657]{Pengfei Jiang}
\affiliation{Xinjiang Astronomical Observatory, CAS, 150 Science-1 Street, Urumqi 830011, China}

\author[0000-0003-4341-0029]{Tao An}
\affiliation{Shanghai Astronomical Observatory, CAS, 80 Nandan Road, Shanghai 200030, China}

\author{Hongmin Cao}
\affiliation{School of Electronic and Electrical Engineering, Shangqiu Normal University, 298 Wenhua Road, Shangqiu, Henan 476000, China}

\author[0000-0002-8684-7303]{Ning Chang}
\affiliation{Xinjiang Astronomical Observatory, CAS, 150 Science-1 Street, Urumqi 830011, China}

\author{Giulia Migliori}
\affiliation{INAF -- Istituto di Radioastronomia, Via Gobetti 101, Bologna 40129, Italy}

\author{Marcello Giroletti}
\affiliation{INAF -- Istituto di Radioastronomia, Via Gobetti 101, Bologna 40129, Italy}

\author[0000-0003-3079-1889]{S\'andor Frey}
\affiliation{Konkoly Observatory, HUN-REN Research Centre for Astronomy and Earth Sciences, Konkoly Thege Mikl\'os \'ut 15-17, H-1121 Budapest, Hungary}
\affiliation{CSFK, MTA Centre of Excellence, Konkoly Thege Mikl\'os \'ut 15-17, H-1121 Budapest, Hungary}
\affiliation{Institute of Physics and Astronomy, ELTE Eötvös Loránd University, P\'azm\'any P\'eter s\'et\'any 1/A, H-1117 Budapest, Hungary}

\author[0000-0002-2322-5232]{Jun Yang}
\affiliation{Department of Space, Earth and Environment, Chalmers University of Technology, Onsala Space Observatory, Onsala 43992, Sweden}

\author[0000-0003-1020-1597]{Krisztina \'E. Gab\'anyi}
\affiliation{Department of Astronomy, Institute of Physics and Astronomy, ELTE Eötvös Loránd University, P\'azm\'any P\'eter s\'et\'any 1/A, H-1117 Budapest, Hungary}
\affiliation{HUN-REN--ELTE Extragalactic Astrophysics Research Group, ELTE E\"otv\"os Lor\'and University, P\'azm\'any P\'eter s\'et\'any 1/A, H-1117 Budapest, Hungary}
\affiliation{Konkoly Observatory, HUN-REN Research Centre for Astronomy and Earth Sciences, Konkoly Thege Mikl\'os \'ut 15-17, H-1121 Budapest, Hungary}
\affiliation{CSFK, MTA Centre of Excellence, Konkoly Thege Mikl\'os \'ut 15-17, H-1121 Budapest, Hungary}

\author{Xiaoyu Hong}
\affiliation{Shanghai Astronomical Observatory, CAS, 80 Nandan Road, Shanghai 200030, China}

\author{Wenda Zhang}
\affiliation{National Astronomical Observatories, CAS, 20A Datun Road, Beijing 100012, China}



\begin{abstract}
Understanding the formation mechanisms of stellar-mass black holes in X-ray binaries (BHXBs) remains a fundamental challenge in astrophysics. The natal kick velocities imparted during black hole formation provide crucial constraints on these formation channels. In this work, we present a new-epoch very long baseline interferometry (VLBI) observation of the Galactic BHXB AT2019wey carried out in 2023. Combining with archival VLBI data from 2020, we successfully measure the proper motion of AT2019wey over a 3-year timescale, namely $0.78\pm0.12$~\masyr\ in right ascension and $-0.42\pm0.07$~\masyr\ in declination. Employing the measured proper motion, we estimate its peculiar velocity and the potential kick velocity (PKV), through Monte Carlo simulations incorporating uncertainties of its distance and radial velocity. The estimated PKV distributions and height above the Galactic plane suggest that AT2019wey's black hole likely formed through a supernova explosion rather than direct collapse.

\end{abstract}



\section{Introduction} \label{sec:intro}

Black hole X-ray binaries (BHXBs) are binary systems consisting of a stellar-mass black hole accreting matter from a companion, typically a low-mass evolved star. BHXBs mostly remain in a quiescent state before transitioning into an outburst \citep{2006ARA&A..44...49R}. During these outbursts, BHXBs exhibit luminous X-ray emissions, but the compact radio jets are often quenched during the soft spectral state. Additionally, the accretion rate of black holes in BHXBs can fluctuate significantly over short timescales \citep{2012Sci...337..540F,2014SSRv..183..323F}. BHXBs serve as critical astrophysical laboratories that enable us to study a wide range of phenomena related to accretion physics, jet formation, strong gravity effects around stellar-mass black holes, and binary evolution \citep{1997ApJ...489..865E,2006csxs.book..157M,2006ARA&A..44...49R,2009MNRAS.396.1370F,2010MNRAS.403...61D,2021Galax...9...78L,2022hxga.book...81A,2023hxga.book..139B}. 

Variations in the accretion flow of BHXBs can significantly change the X-ray spectral energy distribution. The accretion states of BHXBs can be broadly classified into hard and soft states \citep[e.g.][]{2004ARA&A..42..317F,2012Sci...337..540F}. Two distinct states of BHXBs correspond to different behaviours observed at radio wavelengths. The hard state is characterized by a continuous jet, while the soft state has never been observed to be associated with a quasi-steady jet \citep[e.g.][]{2001MNRAS.327.1273S,2011ApJ...739L..19R,2014SSRv..183..323F,2020NatAs...4..697B}.
Very long baseline interferometry (VLBI) is a powerful technique that provides the highest angular resolution for imaging the compact jets and structures in BHXBs \citep{1995Natur.374..141T,1999ARA&A..37..409M,2001MNRAS.327.1273S,2004ARA&A..42..317F,2008JPhCS.131a2057M}. It also enables precise astrometry and measurement of proper motions of BHXBs \citep[e.g.][]{2009ApJ...706L.230M,2011MNRAS.415..306M,2011ApJ...742...83R,2014ApJ...796....2R,2020MNRAS.493L..81A}, providing insights into their formation mechanisms based on their space velocities \citep[e.g.][]{2007ApJ...668..430D,2009MNRAS.394.1440M,Atri2019}.

An example of such a system, which exhibits many of the remarkable properties of BHXBs, is AT2019wey, a newly-identified Galactic microquasar hosting a candidate stellar-mass black hole accretor. Since its initial detection as an optical transient in late 2019 by Asteroid Terrestrial-impact Last Alert System \citep[ATLAS,][]{2019TNSTR...3....1T} and subsequent association with an X-ray source \citep{2020ATel13571....1M}, AT2019wey has been the subject of extensive multi-wavelength campaigns from radio to X-rays \citep{2020ATel13932....1Y,2020ATel13984....1C,2020ATel14168....1G,2021ApJ...909L..27Y,2021ApJ...920..120Y,2021ApJ...920..121Y}. These observations, coupled with the detection of a persistent compact jet in VLBI imaging \citep{2021ApJ...920..121Y, 2022A&A...657A.104C}, firmly established AT2019wey as a newly-identified Galactic microquasar -- a low-mass BHXB system exhibiting relativistic jets. Moreover, early VLBI studies of AT2019wey also resolved a flat-spectrum radio source on milliarcsecond (mas) scales and constrained a lower distance limit of $>6$ kpc based on the angular broadening imposed by interstellar scattering \citep{2021ApJ...909L..27Y,2022A&A...657A.104C}. However, the relatively short VLBI epoch timelines (4 epochs within an interval of $\sim$ 1 month) precluded a direct measurement of the system's proper motion during these initial VLBI observations. 

In this work, we present a new-epoch VLBI observation made in 2023, allowing us to measure the proper motion of AT2019wey by combining with the 2020 observations over a 3-year time baseline. These high-precision astrometric measurements allow us to probe the space velocity of this Galactic BHXB for the first time. Characterizing the jet properties and space motions is crucial for understanding accretion/ejection processes, binary evolution, and the formation of stellar-mass black holes. Measuring proper motions provides unique insights into the intrinsic velocities imparted by possible natal kicks during the supernova (SN) explosion that created the black hole. Our new VLBI observation of AT2019wey provides a unique opportunity to connect its proper motion to the black hole's formation scenario and subsequent kinematic evolution in the Galactic environment. 

\section{VLBI Observation and Data Reduction} \label{sec:data}

Three years after the last multi-wavelength VLBI observations, we carried out again a $\sim$4 h observation of AT2019wey at 4.926~GHz with the European VLBI Network (EVN) on 2023 October 4 (project code EC095). The radio telescopes Effelsberg (Germany), Jodrell Bank MK\uppercase\expandafter{\romannumeral2} (United Kingdom), Westerbork (The Netherlands), Medicina (Italy), Onsala (Sweden), Tianma (China), Toru\'n (Poland), and Irbene (Latvia) were involved in this project. The data were recorded at the rate of 2048~Mbps (8$\times$32~MHz subbands, full polarization, two-bit sampling) and correlated in e-VLBI mode \citep{Szomoru2008}. We utilized the phase-referencing technique to interleave scans between the phase-referencing calibrator J0442+5436 and the target AT2019wey (separated by $1\fdg3$). The cycle time was 4 min, including 3 min for J0442+5436 and 1 min for AT2019wey. A bright and compact source DA193 was inserted to be observed in the middle of these VLBI scans as a fringe-finder. 

We reduced the correlated data using the Astronomical Image Processing System \citep[\textsc{aips},][]{Greisen2003}. First, we flagged data when the antennas were slewing, or had other instrumental or recording problems. Parallactic angle and ionospheric corrections were then performed. A priori amplitude calibration was conducted with gain curves and system temperature measurements of each antenna. After manual phase calibrations on both DA193 and J0442+5436 to remove the instrumental delays, we performed the global fringe fitting on DA193 and J0442+5436, respectively. After that, imaging and self-calibrations in the \textsc{difmap} software \citep{Shepherd1994} were performed on the phase-referencing calibrator J0442+5436. Next, we performed the global fringe fitting in \textsc{aips} again using the obtained image of J0442+5436 as the input model to correct the phase errors caused by the structure of this calibrator. The phase and amplitude self-calibrations were also conducted with the obtained image of J0442+5436 in \textsc{aips} to reduce the residual phase and amplitude errors. Finally, we interpolated the derived solutions to the target AT2019wey in \textsc{aips}, and imaged this target without self-calibration in \textsc{difmap}.

\section{Results} \label{sec:results}

\subsection{The New VLBI Image}

\begin{figure}
\centering
\includegraphics[width=0.6\columnwidth]{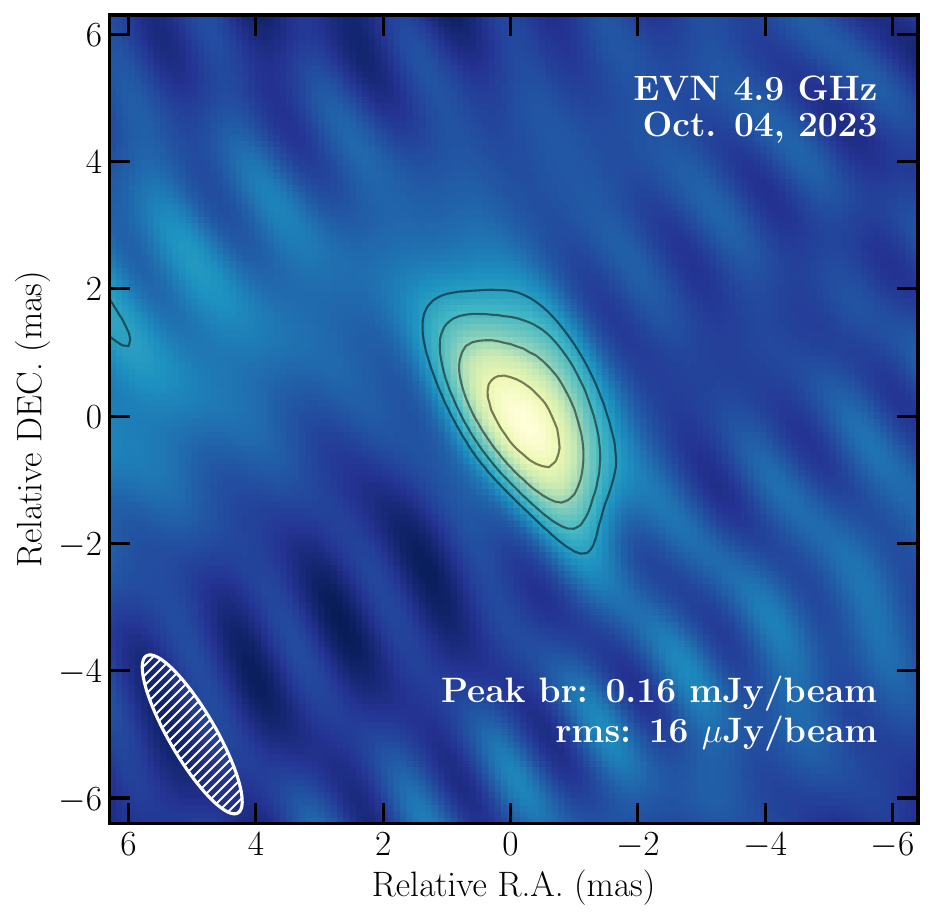}
\caption{Naturally weighted image of AT2019wey obtained with the 4.9 GHz EVN observation on 2023 October 4. The peak brightness value and the image noise level are labeled in the lower right corner. Contours start from 3$\sigma$ noise level and increase by a factor of $\sqrt{2}$. The full width at half maximum (FWHM) of the Gaussian synthesized beam is represented by an ellipse in the lower left corner ($2.86 \times 0.73$~mas$^2$, major axis position angle PA=$30\fdg2$).}
 \label{fig:fig1}
\end{figure}

The naturally weighted image of AT2019wey obtained with the new EVN observation is presented in Figure~\ref{fig:fig1}. The source appears resolved in this image, as it is slightly wider than the elliptical model of the synthesized beam in the southeast--northwest direction. A fit to the calibrated visibility data with a circular Gaussian model gives a total flux density of 0.46$\pm0.16$ mJy and a size of 1.6$\pm0.5$ mas for this source. Here, we used the method provided by \citet{Lee2008} to estimate the uncertainties of the flux density and source size. The fitted size is much larger than the theoretical minimum resolvable size \citep{Lobanov2005} of $\sim$ 0.4 mas in this naturally weighted image, confirming that the radio emission from AT2019wey is resolved in this observation. 

\citet{2022A&A...657A.104C} suggested that the resolved feature of AT2019wey on milliarcsecond scales could be explained as scatter broadening caused by the Galactic interstellar medium, based on its source angular size versus frequency trend. Compared with the results from \citet{2022A&A...657A.104C}, we found a good consistency between our fitted size (1.6 $\pm0.5$ mas at 4.9 GHz) and those derived from the previous two-epoch VLBA observations at a close observing frequency, namely 2.0$\pm0.4$ mas and 1.4$\pm0.3$ mas at 4.5 GHz on November 24 and December 09 in 2020, respectively. Given this, our new result could also support the hypothesis of scatter broadening.

\subsection{Detection of Proper Motion}

High precision in proper motion measurements is more easily achieved over a longer observing timeline, as proper motion is a time-dependent linear change in the source position. Based on multi-epoch VLBI observations spanning $\sim$ 2 months, \citet{2022A&A...657A.104C} set an upper limit on the proper motion of AT2019wey: $\mu_{\alpha}<3.8$\,\masyr\ and $\mu_{\sigma}<0.9$\,\masyr\ in right ascension and declination, respectively. Combined with the new EVN observation, higher precision in proper motion measurements can be attained with a significantly extended observing span of $\sim3$ yr.

To determine the proper motion of AT2019wey, we measured the position of this source at each observation and analyzed the position errors. The \textsc{aips} task \texttt{JMFIT} was used to derive the peak position and the statistical uncertainty for images of each epoch. For phase-referencing observations, astrometric uncertainties are mainly caused by the ambiguity of station coordinates, Earth orientation, and troposphere errors. Based on the simulation of \citet{Pradel2006}, we estimated that the astrometric uncertainties are $\sim 0.07$ mas in right ascension and $\sim 0.08$ mas in declination for EVN observations, while $\sim 0.05$ mas in right ascension and $\sim 0.08$ mas in declination for VLBA observations. Besides that, the astrometric errors caused by the residual unmodelled ionosphere cannot be neglected. Following an analytical calculation \citep{ReidHonma2014,Rioja2020}, our estimated error values are $\sim 0.07$ mas, $\sim 0.06$ mas, and $\sim 0.03$ mas at 4.5 GHz, 4.9 GHz, and 6.7 GHz, respectively. Nevertheless, the position derived from the new EVN observation requires more comments. The participation of the Tianma telescope greatly improved the resolution along the southeast--northwest direction, yet it concurrently led to a poor $(u,v)$ coverage, leaving a significant void for $(u,v)$ radius at $\sim 30-110$ M$\lambda$ ($\lambda$ is the observing wavelength). The asymmetric resolved structure along the southeast--northwest direction and the possible artificial structure caused by the poor $(u,v)$ coverage could make the measured reference point position change, resulting in astrometric errors for the proper motion fitting. Therefore, a more conservative error estimate for this position measurement should be provided. Given this, we measured the position of AT2019wey without Tianma data, and the differences between the position measurements are 0.34 mas in right ascension and 0.06 mas in declination. These are probably the largest potential position errors arising from the target source structure issues in this observation. For the phase-referencing calibrator J0442+5436, the same coordinates were used throughout all observations, thereby no additional error was propagated to the relative position measurements of the target source. We also examined the discrepancies in position when utilizing the structural model of the phase-referencing calibrator derived from different observing epochs for the new EVN observation, as the coverage of the $(u,v)$ plane of the new observation is obviously different from the other observations. The error in the right ascension direction is at a negligible level of 0.01 mas, while in the declination direction is smaller than $\sim$ 0.1 mas. Finally, all errors are propagated to the source positions and summarized in Table \ref{tab:tab1}.

\begin{figure}
\centering
\includegraphics[width=\columnwidth]{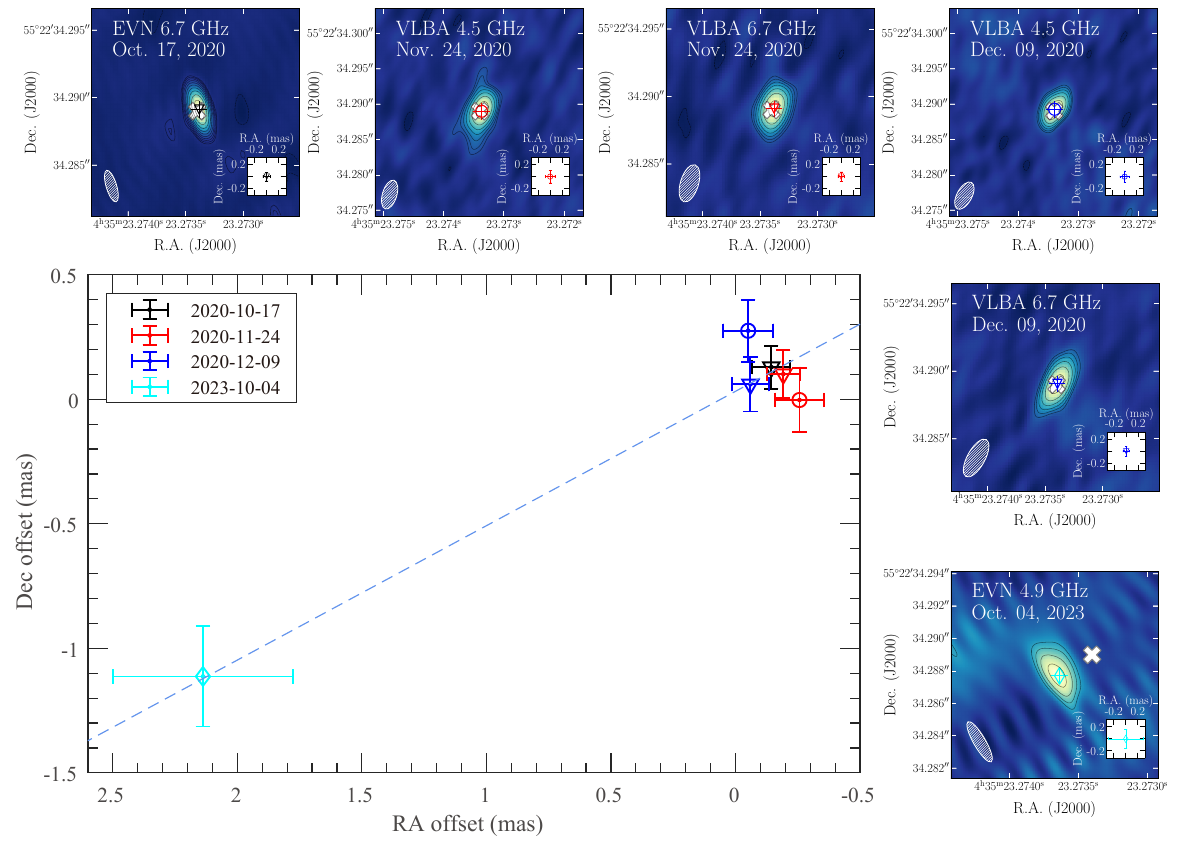}
\caption{The sky position evolution of AT2019wey over 2020 to 2023 derived from multi-epoch VLBI observations is shown in the bottom left panel. The blue dashed line represents the fitting proper motion. The surrounding panels are VLBI images of all epochs, while the colored markers represent the fitted positions listed in Table 1, with the error shown at the lower right corner in each sub-figure. The colors and shapes of the markers in the bottom left panel indicate different observation epochs and frequencies: triangles represent 6.7 GHz, circles represent 4.5 GHz, and a diamond represents the new 4.9 GHz. 
The synthesized beam sizes are shown in the bottom left corner of these panels.}
\label{fig:fig2}
\end{figure}

We applied a least-squares linear regression on the position time series to fit the proper motion of AT2019wey. As shown in Figure~\ref{fig:fig2}, the measured proper motion is $\mu_{\alpha}=0.78\pm0.12$\,\masyr\ and $\mu_{\delta}=-0.42\pm0.07$\,\masyr\ . The reduced $\chi^2$ per
degree of freedom of $\sim$ 0.9 is near unity, implying that the position errors are only slightly overestimated. Compared with the results from \citet{2022A&A...657A.104C}, our new values of proper motion fall well within the previous upper limit range but provide more precise estimates.

\begin{table*}
    \centering
    \caption{Phase-referenced coordinates of AT2019wey derived from \texttt{JMFIT}.}
    \label{tab:tab1}
    \begin{tabular}{ccccccc}
        \hline
        Epoch & Obs.data & Frequency & $\alpha$ & $\sigma_{\alpha\cos\delta}$ & $\delta$  & $\sigma_{\delta}$  \\
              & (y m d)  &(GHz)      & (h m s)  & (mas)                       & (d m s)   & (mas)              \\
        \hline
        1       &2020-10-17 &6.7              &04 35 23.273383 &0.08 &55 22 34.28913 &0.09  \\[2pt]
        2       &2020-11-24 &4.5              &04 35 23.273370 &0.10 &55 22 34.28900 &0.13  \\[2pt]
                &           &6.7              &04 35 23.273378 &0.07 &55 22 34.28910 &0.10  \\[2pt]
        3       &2020-12-09 &4.5              &04 35 23.273394 &0.10 &55 22 34.28928 &0.12  \\[2pt]
                &           &6.7              &04 35 23.273393 &0.07 &55 22 34.28906 &0.11  \\[2pt]
        4       &2023-10-04 &4.9              &04 35 23.273651 &0.36 &55 22 34.28789 &0.20  \\[2pt]
        
        \hline
    \end{tabular}\\
\end{table*}

\subsection{Peculiar Velocity and Potential Kick Velocity}
\label{subsec:section3.3}

If a celestial body's position, distance, radial velocity, and proper motion are known, its peculiar velocity could also be estimated based on a model of Galactic rotation \citep[e.g.][]{Reid2009}. In this paper, we provided the position and proper motion information for AT2019wey. For the distance of this source, \citet{2021ApJ...920..120Y} set an upper limit of 10 kpc and \citet{2022A&A...657A.104C} provided a lower limit of 6 kpc. However, the information on the radial velocity is lacking for this source. Here we estimate the expected radial velocity $v_{\rm rad}$ of AT2019wey assuming the system undergoing pure Galactic rotation around the Galactic center \citep[e.g.][]{Atri2019}. We followed the calculation method from \citet{Reid2009}, adopting solar motion parameters of ($U_{\odot}$ = 10.7$\pm$1.8, $V_{\odot}$ = 15.6$\pm$6.8, $W_{\odot}$ = 8.9$\pm$0.9)\,\kms\ \citep{Reid2014} and a Galactic model where the Sun is assumed to be at 8 kpc from the Galactic center and the circular velocity is 220\,\kms\  for a flat rotation curve \citep[e.g.][]{Bovy2012}. At distances of [6, 8, 10] kpc, the derived radial velocities are [$-43$, $-51$, $-57$]\,\kms\,. With those derived radial velocities, the corresponding peculiar velocities are [27, 37, 46]\,\kms, respectively.

\begin{figure*}
	\centering
	\begin{minipage}[b]{0.48\textwidth}
		\centering
		\includegraphics[width=\linewidth]{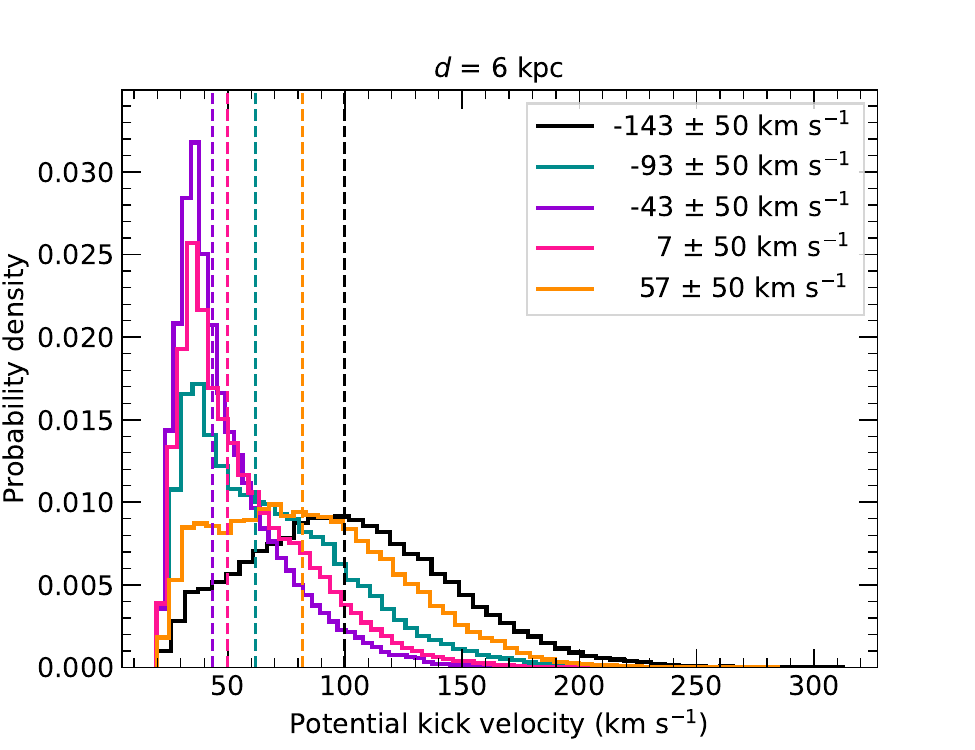}
	\end{minipage}%
        \\
	\begin{minipage}[b]{0.48\textwidth}
		\centering
		\includegraphics[width=\linewidth]{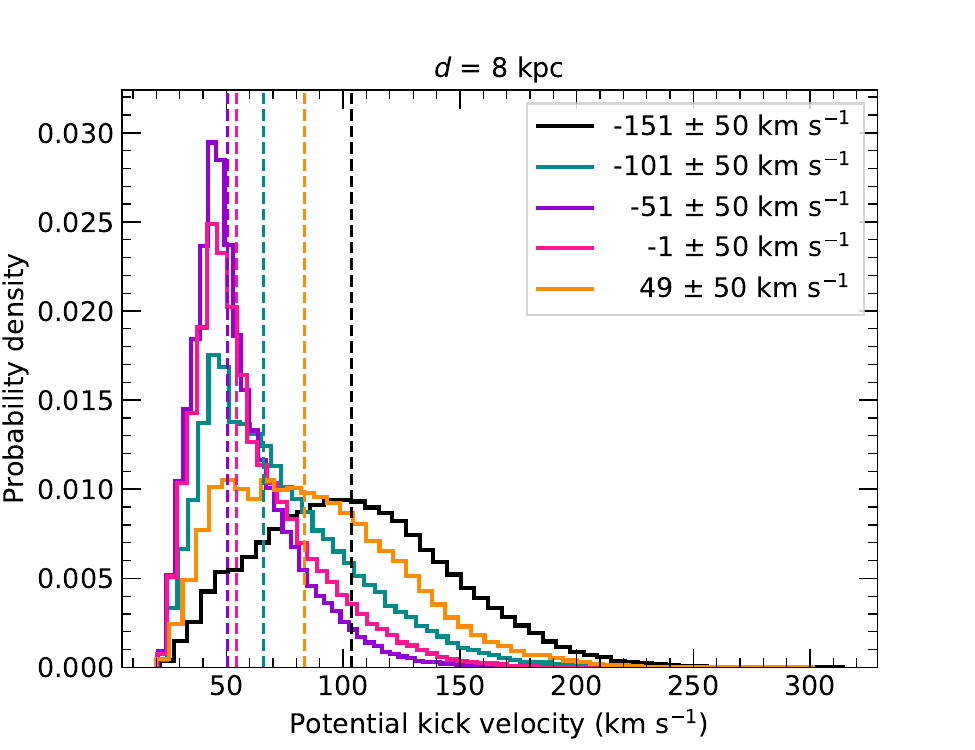}
	\end{minipage}%
        \\
	\begin{minipage}[b]{0.48\textwidth}
		\centering
		\includegraphics[width=\linewidth]{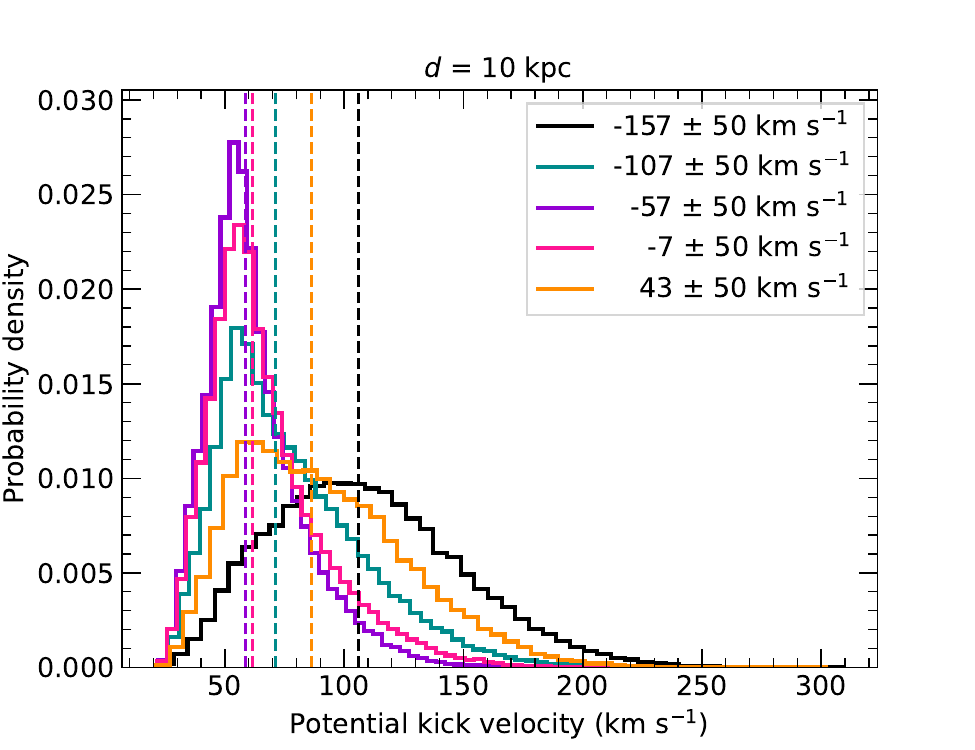}
	\end{minipage}%
	\caption{PKV probability distributions of AT2019wey using five probable Gaussian radial velocity distributions (noted in the legend of each panel) at three distances of [6, 8, 10] kpc. The vertical dashed lines represent the medians of the corresponding distributions. The parameters are listed in Table~\ref{tab:tab2}}.
	\label{fig:fig4}
\end{figure*}

Furthermore, we estimate the potential kick velocity (PKV) for AT2019wey, which represents the peculiar velocity of the system when it crosses the Galactic plane \citep{Atri2019}. AT2019wey is not located precisely in the Galactic plane. Its galactic latitude of $5.3\degr$ and a distance lower limit of 6 kpc \citep{2022A&A...657A.104C} imply that its height above the Galactic plane is greater than $\sim$550 pc. Given that the majority of star-forming regions reside within the Galactic plane, there is a high probability that AT2019wey was also formed within the plane. Note that there is still a possibility that the system could have been formed in, and ejected from, a globular cluster. Without better information on the age and nature of the donor, or about the radial velocity and distance and hence the exact orbit of the system, it is difficult to robustly rule out this possibility currently. In the case of being formed in the Galactic plane, AT2019wey is unlikely to be undergoing pure Galactic rotation about the Galactic center, since it should at least have some component of $z$-velocity (the velocity along the direction perpendicular to the Galactic plane), oscillating above and below the Galactic plane. In such an orbit, the peculiar velocity is not a conserved quantity. Hence PKV, the peculiar velocity of a system when it crosses the Galactic plane, serves as a more effective tool to comprehend the natal kick a BHXB received when the BH was born. Additionally, it should be noted that the estimated radial velocities do not accurately reflect the true radial velocity of the source, as it may have undergone kicks, thereby likely not adhering to a pure Galactic rotation about the Galactic center. Therefore, we adopted five possible Gaussian radial velocity distributions with medians of $v_{\rm rad}$, $v_{\rm rad}\pm50$\,\kms, and $v_{\rm rad}\pm100$\,\kms\, in the following calculations. Here we use the same code\footnote{\url{https://github.com/pikkyatri/BHnatalkicks}} reported in \citet{Atri2019} incorporating the Monte Carlo (MC) simulation technique to estimate the PKV probability distribution for AT2019wey. The PKV probability distributions of AT2019wey using five probable Gaussian radial velocity distributions at three distances of [6, 8, 10] kpc are plotted in Figure~\ref{fig:fig4} and their parameters are given in Table~\ref{tab:tab2}. 

\begin{table}
    \centering
    \caption{PKV probability distribution of AT2019wey using five probable Gaussian radial velocity distributions at three distances of [6, 8, 10] kpc. The PKV value given here represents the median of PKV probability distribution, with the lower and upper limits corresponding to the 5th and 95th percentiles, respectively.}
    \label{tab:tab2}
    \begin{tabular}{ccc}
        \hline
        $d$ & $v_{\rm rad}$ & PKV  \\
        (kpc)  &(\kms)      & (\kms) \\
        \hline
         6 & $-143\pm50$ & $100_{-62}^{+78}$ \\[2pt]
           & \hspace{4pt}$-93\pm50$ & \hspace{4pt}$62_{-34}^{+71}$ \\[2pt]
           & \hspace{4pt}$-43\pm50$ & \hspace{4pt}$44_{-18}^{+54}$ \\[2pt]
           & \hspace{16pt}$7\pm50$ & \hspace{4pt}$50_{-23}^{+62}$ \\[2pt]
           & \hspace{12pt}$57\pm50$ & \hspace{4pt}$82_{-50}^{+72}$ \\[2pt]
        \hline
         8 & $-151\pm50$ & $103_{-58}^{+76}$ \\[2pt]
           & $-101\pm50$ & \hspace{4pt}$66_{-32}^{+70}$ \\[2pt]
           & \hspace{4pt}$-51\pm50$ & \hspace{4pt}$51_{-20}^{+48}$ \\[2pt]
           & \hspace{9pt}$-1\pm50$ & \hspace{4pt}$54_{-22}^{+59}$ \\[2pt]
           & \hspace{12pt}$49\pm50$ & \hspace{4pt}$83_{-45}^{+72}$ \\[2pt]
        \hline
         10 & $-157\pm50$ & $106_{-55}^{+75}$ \\[2pt]
           & $-107\pm50$ &  \hspace{4pt}$71_{-32}^{+66}$\\[2pt]
           & \hspace{4pt}$-57\pm50$ & \hspace{4pt}$59_{-23}^{+43}$ \\[2pt]
           & \hspace{9pt}$-7\pm50$ & \hspace{4pt}$62_{-25}^{+55}$ \\[2pt]
           & \hspace{12pt}$43\pm50$ & \hspace{4pt}$86_{-42}^{+71}$ \\[2pt]
        \hline
    \end{tabular}\\
\end{table}

\section{Discussion} \label{sec:disc}

The astrometric results derived from the EVN and VLBA observations spanning $\sim$ 3 years show that AT2019wey has a significant proper motion. Based on the results, we estimate the peculiar velocity and PKV value of this source, which aid us in understanding the kick that the BHXB system received when the BH was born.

Given the distance uncertainty of this system and its unknown radial velocity, we provide the estimated peculiar velocity and the probability distributions of the PKV value under multiple scenarios in Section~\ref{subsec:section3.3}. With the assumption that the expected radial velocity is determined by the pure Galactic rotation at the Galactic longitude and distance of AT2019wey, and without considering its dispersion, the estimated peculiar velocity provides a relatively small value relative to PKV. PKV, the peculiar velocity of a system at Galactic plane crossing, is a better proxy to investigate the BH natal kick. For a distance of 6 kpc and the central predicted radial velocity, the PKV probability distribution exhibits a significant positive skew, with a median value of $\sim$ 40\,\kms\ and the highest probability density at $\sim$ 30\,\kms\ . These values are lower than the typical velocity dispersion of stars in the Galaxy ($50 \, \mathrm{km} \, \mathrm{s}^{-1}$; \citealt{Mignard2000}), which suggests that AT2019wey is likely to have formed either in a supernova (SN) with a low natal kick or via direct collapse. However, the median PKV exceeds $\sim$ 50\,\kms\ for the other four predicted radial velocities. Furthermore, all median PKV values fall within the range of $\sim$ 50 -- 110\,\kms\ for the distance of 8 and 10 kpc. 

Based on 3D supernova simulations and semi-analytical parametrized models, \citet{Mandel2020} developed a probabilistic model for natal kick velocities for neutron stars and BHs. For the latter, the probability distribution indicates that most BHs receive no natal kick at all, while the non-zero velocities can reach up to $\sim150$~km\,s$^{-1}$, with typical values around $60$~km\,s$^{-1}$. Our estimated median PKV values of AT2019wey are consistent with this theoretical picture. Since BHXBs are more massive than common systems, they will suffer even lower velocity dispersions than that of common stars ($50 \, \mathrm{km} \, \mathrm{s}^{-1}$; \citealt{Mignard2000}). In this case, these median PKV values fall within the range of $\sim$ 50 -- 110\,\kms\ suggesting that AT2019wey is likely to be formed by an SN explosion rather than direct collapse. 

The height of BHXB systems above the Galactic plane could be a supplementary proxy for the BH natal kick \citep[e.g.][]{Jonker2004}. The height above the Galactic plane for AT2019wey is greater than $\sim$ 550 pc. Similar situation is seen in many BHXB systems (the height greater than 500 pc) such as MAXI J1820+070, SAX J1819.3-2525, MAXI J1836-194, GX 339-4, and 4U 1543-475, where their PKVs fall within the range of $\sim$ 60 -- 240\,\kms\ and suggest that these systems have received a strong kick \citep{Atri2019}. Thus the height above the Galactic plane for AT2019wey also indicates that this system may have received a strong kick when the BH was born. 

However, do note that the uncertainty in the distance of AT2019wey and its unknown radial velocity (especially the latter), make it extremely difficult to accurately estimate its PKV value, which in turn significantly complicates deducing its likely birth mechanism. Additionally, \citet{Nagarajan2024} suggested that PKV is likely to overestimate the kick velocity, as it does not take into account the velocity dispersion of the local stellar population. For example, they found that MAXI J1820+070 needed a minimum kick of only 14 \,\kms, and SAX J1819.3-2525 a minimum kick of only 44\,\kms. \citet{Nagarajan2024} argued that, after accounting for the stellar velocity dispersion, the kicks required for many of the known Galactic BHs seem to be overestimated in the literature. In our case, the distance of AT2019wey is uncertain, thus a study similar to that of \citet{Nagarajan2024} is not feasible at present to check if the kick for AT2019wey is also overestimated. Nevertheless, this underlines the difficulties and also the importance of accurate determination of natal kick velocities for Galactic BHs. We therefore encourage follow-up observations on AT2019wey to obtain more precise measurements or constraints on its distance and radial velocity, which are still crucial for fully reconstructing its three-dimensional motion and formation history. 


\begin{acknowledgments}
We thank the referee for the insightful and detailed comments which
greatly helped us improve the manuscript. The European VLBI Network (EVN) is a joint facility of independent European, African, Asian, and North American radio astronomy institutes. Scientific results from data presented in this work are derived from the EVN project code: EC095.
This work was supported by the CAS `Light of West China' Program (grant No. 2021-XBQNXZ-005), the National SKA Program of China (grant No. 2022SKA0120102). L.C. acknowledges the support from the Tianshan Talent Training Program (grant No. 2023TSYCCX0099). T.A. and N.C. acknowledge the support from the Xinjiang Tianchi Talent Program. T.A. is supported by FAST Special Grant of NSFC (12041301). This work was also partly supported by the Urumqi Nanshan Astronomy and Deep Space Exploration Observation and Research Station of Xinjiang (XJYWZ2303). 
S.F. and K.\'E.G. thank the Hungarian National Research, Development and Innovation Office (NKFIH) for support (OTKA K134213). Their work was also supported by HUN-REN and the NKFIH excellence grant TKP2021-NKTA-64.

\end{acknowledgments}

\bibliography{sample631}{}
\bibliographystyle{aasjournal}



\end{document}